\documentclass[10pt,twocolumn,letterpaper]{article}

\usepackage[pagenumbers]{cvpr}

\def\paperID{*****}
\def\confName{CVPR}
\def\confYear{2026}

\usepackage{microtype}
\usepackage{algorithm}
\usepackage{algorithmic}
\definecolor{cvprblue}{rgb}{0.21,0.49,0.74}
\usepackage[pagebackref,breaklinks,colorlinks,allcolors=cvprblue]{hyperref}

\newcommand{\R}{\mathbb{R}}
\newcommand{\cM}{\mathcal{M}}
\newcommand{\cL}{\mathcal{L}}
\newcommand{\norm}[1]{\lVert #1 \rVert}
\newcommand{\inv}[1]{\textcolor{black!45}{#1}}

\title{Continuous Neural Reparameterization as a Deep Geometric Prior for Robust Fixed-Chart UV Repair}
\author{Mohammad Sadegh Salehi\\
Zero One Creative\\
London, UK\\
{\tt\small sadegh@01c.ai}
}

\begin{document}

\maketitle

\begin{figure*}[t]
    \centering
    \includegraphics[width=\textwidth]{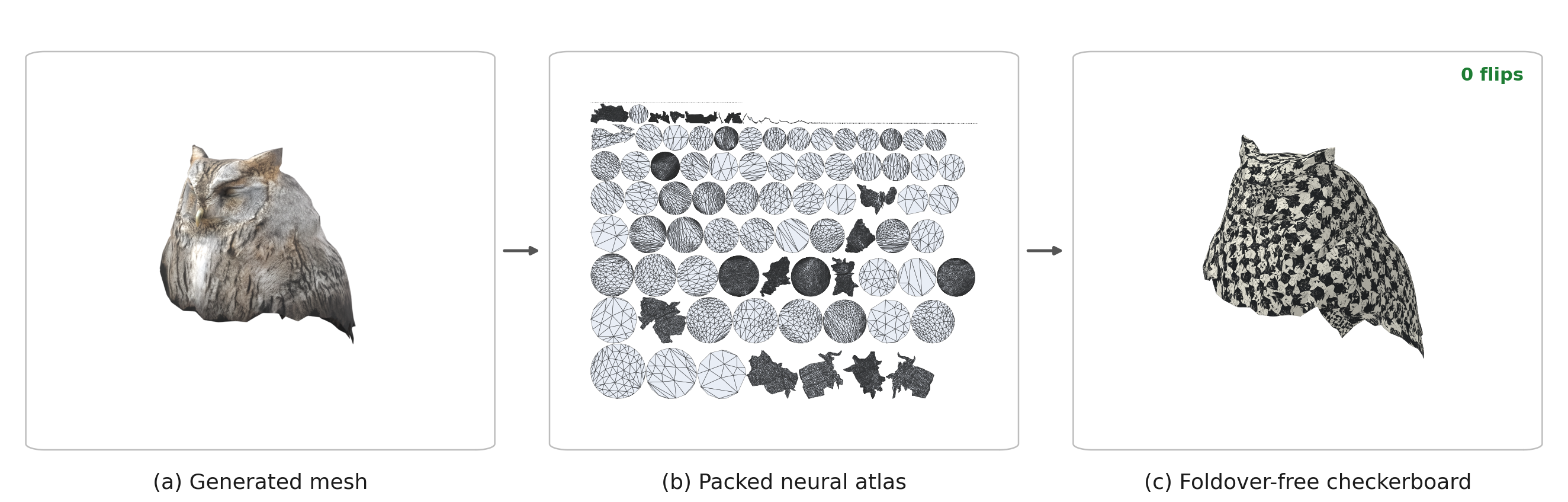}
    \caption{We recast fixed-chart UV unwrapping as continuous neural reparameterization and embed it in validation-first atlas construction. Given a generated mesh (a), chart candidates that pass disk-like validation are flattened by an untrained SIREN optimized for a determinant-aware distortion objective; remaining fragments are routed to marked fallbacks and the result is packed into an atlas (b). A checkerboard transfer through the atlas (c) is foldover-free, with zero flipped triangles. Asset from the Amara Spatial dataset.}
    \label{fig:teaser}
\end{figure*}

\begin{abstract}
Traditional UV unwrapping relies on direct optimization of geometric distortion energies and can fail through invalid initialization, local minima, or topological foldovers. We recast fixed-chart UV unwrapping as continuous neural reparameterization: an untrained SIREN maps per-vertex mesh features to UV coordinates, and its weights are optimized for a geometric objective. The practical contribution is a robust chart-solver recipe, combining Laplace--Beltrami spectral inputs, Tutte residual warm-up, a $C^2$ determinant extension, an injectivity barrier, and validity-checked retry/fallback routing, rather than a claim that any single component guarantees validity or that recutting methods should be replaced. NTK--LBO diagnostics show that spectral conditioning changes update geometry, especially at initialization and mid-rank subspaces, but does not by itself predict chart success. On compact pre-cut charts and a 47-chart stratified Thingi10K/xatlas-cut benchmark, the neural solver produces zero flips on all compact charts and 42/47 valid zero-flip stratified solves. BFF and OptCuts comparisons sharpen the scope: recutting can be faster and lower-distortion when allowed, while the neural solver targets supplied-chart validity and validation-first atlas construction. On Amara Spatial generated meshes, the full atlas construction path gives packed-atlas coverage on a 25-asset set and 1{,}000/1{,}000 strict locally valid atlases with zero UV flips in a large-scale Rust atlas run after fallback routing.
\end{abstract}


\section{Introduction}
\label{sec:introduction}

\begin{figure}[t]
    \centering
    \includegraphics[width=\linewidth]{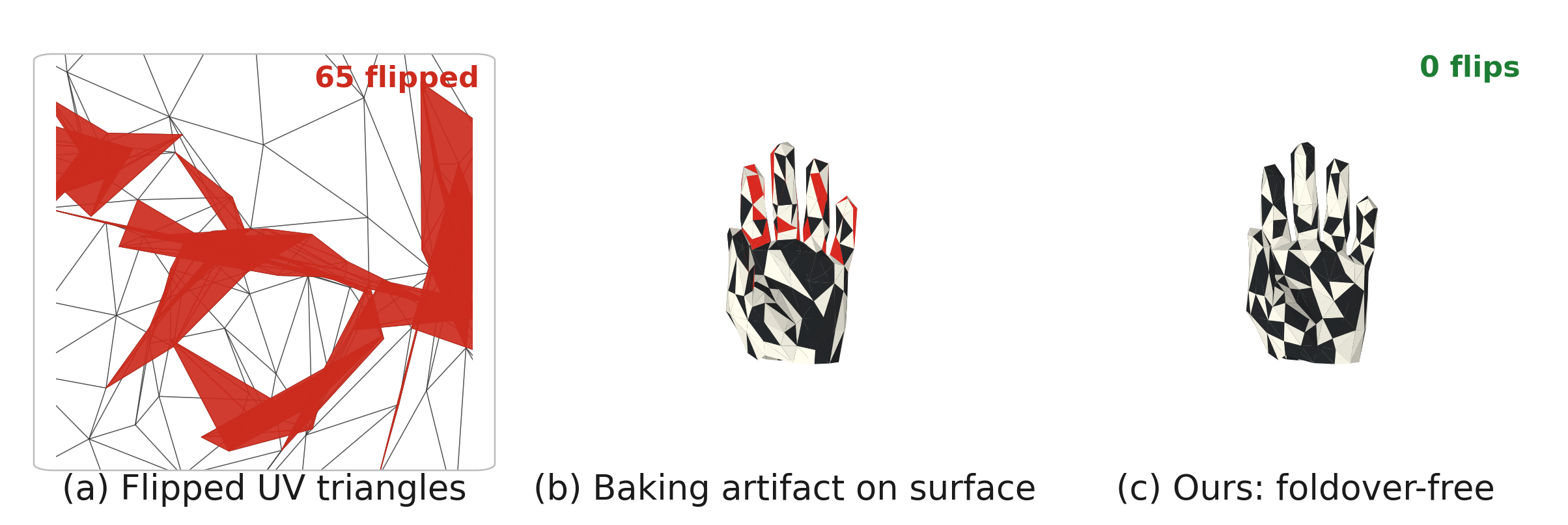}
    \caption{Why local validity matters. (a) A direct solve can fold triangles in the UV plane; the overlapping region (red) contains 65 flipped triangles on this Hand chart. (b) Transferring a texture through that map bakes the foldover into the surface as a visible artifact. (c) Our foldover-free map removes the artifact. The neural solver targets exactly this property: a supplied chart that stays locally valid.}
    \label{fig:motivation}
\end{figure}

Surface parameterization maps a 3D manifold to a 2D domain and is a core operation in texture mapping, remeshing, correspondence, and asset preparation. In UV authoring workflows, the difficult part is not only reducing distortion but preserving local validity: a single flipped UV triangle can invalidate downstream texture baking or require manual repair (Fig.~\ref{fig:motivation}). Classical solvers optimize UV coordinates directly under energies such as the Symmetric Dirichlet energy~\cite{smith2015bijective,rabinovich2017scalable}. These methods are fast and mature, but their optimization landscape can still produce local minima, invalid initialization, or foldovers on difficult charts.

We propose a different parameterization of the unknown map. Instead of treating every UV coordinate as an independent variable, we represent the flattening as the output of an untrained SIREN network~\cite{sitzmann2020implicit} conditioned on mesh coordinates and Laplace--Beltrami eigenfunctions. The network weights are optimized for a geometric UV objective. This turns UV unwrapping into a continuous neural reparameterization problem in which architecture and features provide a useful prior, while local validity is maintained by an explicit recipe: fit a valid Tutte map, optimize a determinant-aware objective, reject invalid steps, and route unsuitable generated fragments to marked fallbacks (Fig.~\ref{fig:teaser}). Our empirical claim is deliberately narrow: the neural solver is useful when a supplied chart must remain locally valid or when cleanup/initialization fails, but it is slower and unnecessary on easy charts, and joint recutting methods can be better when changing seam topology is allowed.

\paragraph{Scope.}
We focus on the chart parameterization subproblem: given a disk-like chart or a collection of validated charts, compute low-distortion UVs with no flipped triangles. Automatic seam placement, chart repair, and packing are evaluated as atlas-construction components, but our central contribution is the neural chart solver.

\paragraph{Contributions.}
\begin{enumerate}
    \item We formulate a robust fixed-chart neural solver combining a moderate-bandwidth SIREN, Laplace--Beltrami spectral features, Tutte residual warm-up, a $C^2$-extended Symmetric Dirichlet objective, determinant-aware rejection, and marked fallback routing.
    \item We provide a failure-mode analysis against tuned SLIM, BFF, and OptCuts, showing where fixed-chart neural optimization helps and where classical conformal or recutting methods are faster, lower-distortion, or better suited.
    \item We report compact chart, Thingi10K/xatlas-cut, and Amara Spatial generated-mesh experiments with explicit success denominators, plus NTK--LBO diagnostics characterizing how spectral conditioning reshapes update subspaces.
\end{enumerate}

\section{Related Work}
\label{sec:related_work}

\paragraph{Mesh parameterization.}
Conformal methods such as LSCM~\cite{levy2002least} and ABF++~\cite{sheffer2005abf++} are efficient but do not guarantee injectivity. ARAP parameterization~\cite{liu2008local} improves local rigidity but can still fold. The Symmetric Dirichlet energy~\cite{smith2015bijective} penalizes both stretch and compression, and SLIM~\cite{rabinovich2017scalable} remains a strong academic baseline through local-global optimization. Boundary-first Flattening~\cite{sawhney2017boundary}, OptCuts~\cite{li2018optcuts}, xatlas~\cite{xatlas}, and Blender~\cite{blender} address chart construction or atlas-generation workflows; these systems complement our chart-level neural solver.

\paragraph{Neural geometric priors.}
Coordinate networks have become standard continuous representations for images, radiance fields, and 3D shape~\cite{mildenhall2020nerf,park2019deepsdf,sitzmann2020implicit}. Fourier features and sinusoidal activations improve high-frequency signal fitting~\cite{tancik2020fourier}; in our setting, the SIREN frequency is a capacity knob rather than a supervised encoder. The Deep Image Prior~\cite{ulyanov2018deep}, Deep Geometric Prior~\cite{williams2019deep}, and Point2Mesh~\cite{hanocka2020point2mesh} show that untrained architectures can regularize inverse problems without external data. Neural Jacobian Fields~\cite{aigerman2022neural} learn intrinsic Jacobian fields from paired mappings and recover maps through a Poisson solve, while Nuvo~\cite{srinivasan2024nuvo} learns neural UV maps for unruly 3D representations. In contrast, we optimize an untrained per-chart UV map directly for local validity and distortion.

\section{Method}
\label{sec:methodology}

Let $\cM=(V,T)$ be a triangular chart. A UV parameterization assigns $\mathbf{u}_i\in\R^2$ to each vertex and induces a per-triangle Jacobian $J_t\in\R^{2\times2}$ once each 3D triangle is expressed in a local tangent basis. This square local representation makes the inverse term below well defined whenever $\det J_t>0$. We measure distortion with
\begin{equation}
    E_{SD}=\sum_{t\in T} A_t\left(\norm{J_t}_{\mathrm{F}}^2 + \norm{J_t^{-1}}_{\mathrm{F}}^2\right),
    \label{eq:esd_short}
\end{equation}
where $A_t$ is the 3D triangle area. Direct methods optimize all $\mathbf{u}_i$ independently. We instead distinguish the induced mesh map $\phi_\theta$ from the coordinate network $f_\theta$ and set
\begin{equation}
    \phi_\theta(v_i)=f_\theta(\mathbf{x}_i), \qquad
    \mathbf{x}_i=[\tilde{\mathbf{p}}_i\|\tilde{\psi}_1(v_i),\dots,\tilde{\psi}_k(v_i)],
\end{equation}
where $\tilde{\mathbf{p}}_i$ is the normalized 3D position and $\tilde{\psi}_j$ are the first $k=16$ non-trivial Laplace--Beltrami eigenfunctions. The SIREN network $f_\theta$ uses sinusoidal activations
\begin{equation}
    \mathbf{h}_{\ell+1}=\sin(\omega_0(W_\ell\mathbf{h}_\ell+\mathbf{b}_\ell)),
\end{equation}
with $\omega_0=15$, five layers, hidden width 256, and a two-dimensional linear output in the reported compact runs. Spectral features provide global chart coordinates, while the finite SIREN frequency range fixes a moderate capacity scale during optimization.

For each triangle $t=(i,j,k)$, we cache local tangent-plane coordinates $\mathbf{c}_i,\mathbf{c}_j,\mathbf{c}_k\in\R^2$ and assemble the affine chart Jacobian as
\begin{equation}
    J_t = [\mathbf{u}_j-\mathbf{u}_i\mid\mathbf{u}_k-\mathbf{u}_i]
    [\mathbf{c}_j-\mathbf{c}_i\mid\mathbf{c}_k-\mathbf{c}_i]^{-1}
    = \Delta U_t\Delta C_t^{-1}.
\end{equation}
This is the discrete Jacobian used for the distortion, determinant barrier, step rejection, and all reported validity metrics.

The neural prior is therefore practical rather than monolithic. LBO features expose intrinsic chart coordinates to the network, gradient descent follows the resulting NTK update geometry, and the valid Tutte warm-up plus determinant barrier discourage leaving the locally injective basin. The SIREN frequency $\omega_0$ remains useful as a capacity setting, but the ablations in Sec.~\ref{sec:experiments} show that it is not the dominant validity factor on the compact charts.

\paragraph{Stable objective.}
The inverse term in Eq.~\eqref{eq:esd_short} is singular when $\det J_t\le 0$. We use a $C^2$ Taylor extension below $\epsilon_J=10^{-4}$ and add an explicit injectivity barrier,
\begin{equation}
    \cL(\theta)=\widetilde{E}_{SD}(\phi_\theta)
    +\alpha(s)\sum_{t\in T}[\epsilon-\det J_t]_+^2,
\end{equation}
with margin $\epsilon=10^{-3}$ and an exponentially annealed weight $\alpha(s)$ from $10$ to $0.1$. Since for $2\times2$ Jacobians $\norm{J_t^{-1}}_{\mathrm{F}}^2=\norm{J_t}_{\mathrm{F}}^2/(\det J_t)^2$, we replace only the singular inverse-determinant factor. For $d=\det J_t\le\epsilon_J$, the factor $d^{-2}$ is replaced by $\epsilon_J^{-2}-2(d-\epsilon_J)\epsilon_J^{-3}+3(d-\epsilon_J)^2\epsilon_J^{-4}$, matching the value, first derivative, and second derivative at $\epsilon_J$. We initialize with a Tutte embedding~\cite{tutte1963draw} and first fit the network to that valid map before optimizing the distortion objective with Adam~\cite{kingma2014adam}.

\paragraph{Spectral-bias diagnostic.}
Motivated by spectral-bias observations in neural networks~\cite{rahaman2019spectral,basri2020frequency}, we treat the network as a prior rather than a learned model. To test whether its training dynamics align with intrinsic mesh harmonics, we compute a subsampled neural tangent kernel (NTK)~\cite{jacot2018neural} $\Theta$ over 256 farthest-point vertices and compare its dominant eigenspace with restricted LBO eigenfunctions. For two $r$-dimensional orthonormal bases $A$ and $B$, we report
\begin{equation}
    s_r=\frac{1}{r}\norm{A^\top B}_{\mathrm{F}}^2 .
\end{equation}
Higher scores indicate that gradient descent updates overlap more strongly with the first intrinsic modes. We use this quantity as a mechanistic diagnostic for how spectral conditioning changes optimization geometry, not as a criterion for predicting whether a chart will succeed.

\section{Implementation}
\label{sec:implementation}

Meshes are normalized to the unit box, repaired by removing degenerate faces and unreferenced vertices, and decomposed into connected chart candidates. Spatial coordinates and each retained eigenfunction are scaled to comparable variance before entering the SIREN, so the first-layer bandwidth does not accidentally privilege one feature family. We compute cotangent LBO eigenfunctions with a lumped mass matrix~\cite{meyer2003discrete} and fix eigenfunction signs deterministically by enforcing positive correlation with the corresponding spatial coordinate when possible, a practical convention for spectral-basis sign ambiguity~\cite{lim2023sign}. The compact-chart solver uses a 5-layer SIREN with width 256, 200 Tutte warm-up iterations, 5{,}000 Adam iterations, cosine learning-rate decay from $5\times10^{-4}$ to $10^{-6}$, determinant-aware step rejection, and early stopping for the efficiency variant. Generated charts use the same solver with a short preview pass and a stricter quality retry schedule. Rejected generated-mesh fragments are explicitly marked as fallback charts using PCA projection or per-face local charts, then all accepted neural and fallback charts are packed with an area-scaled shelf packer. The primary research implementation is in PyTorch; classical compact-chart baselines use libigl 2.6.2, generated-mesh baselines use Blender 4.0.2 loop-level UVs, and a separate Rust atlas implementation is used only for large-scale validation, routing, and packing checks in Sec.~\ref{sec:conclusion}.

\begin{figure*}[!t]
    \centering
    \includegraphics[width=\textwidth]{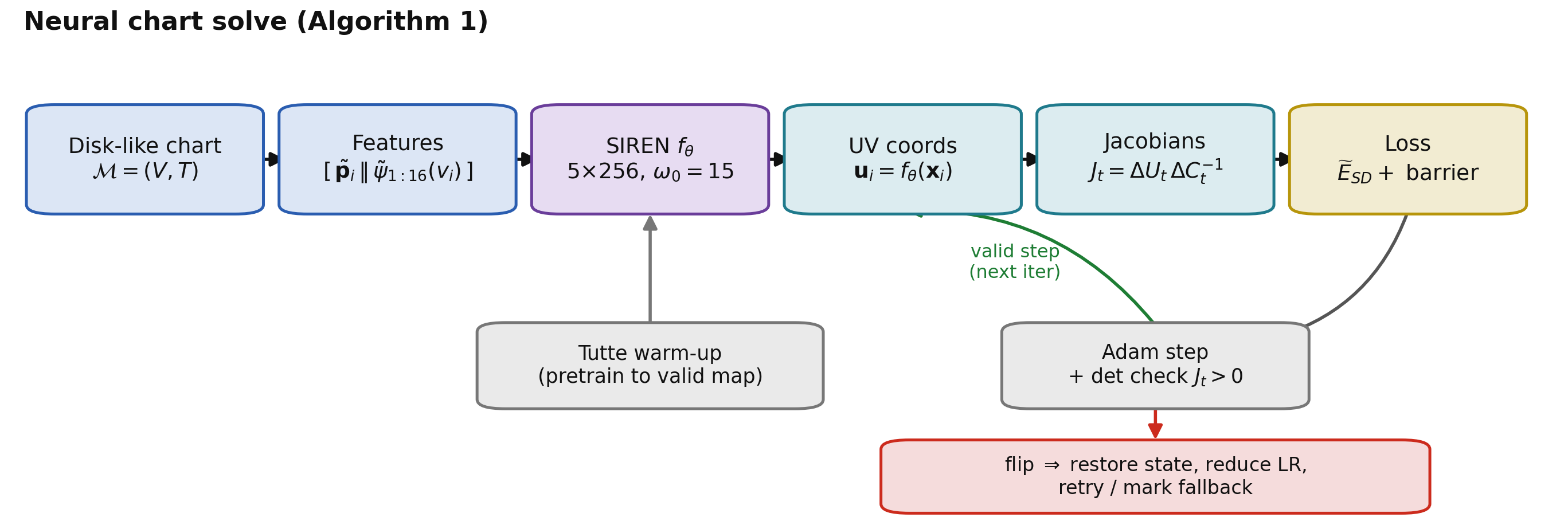}
    \caption{Neural chart solver (Algorithm~\ref{alg:pipeline}). Per-vertex features, normalized position concatenated with $k{=}16$ Laplace--Beltrami eigenfunctions, are mapped by an untrained SIREN to UV coordinates, from which per-triangle Jacobians and a stable Symmetric Dirichlet plus injectivity-barrier loss are assembled. A Tutte warm-up pretrains the network to a valid map; each Adam step is kept only if it remains foldover-free, otherwise the state is restored and the chart is retried or marked as a fallback.}
    \label{fig:solver_schematic}
\end{figure*}

\begin{figure*}[!t]
    \centering
    \includegraphics[width=0.92\textwidth]{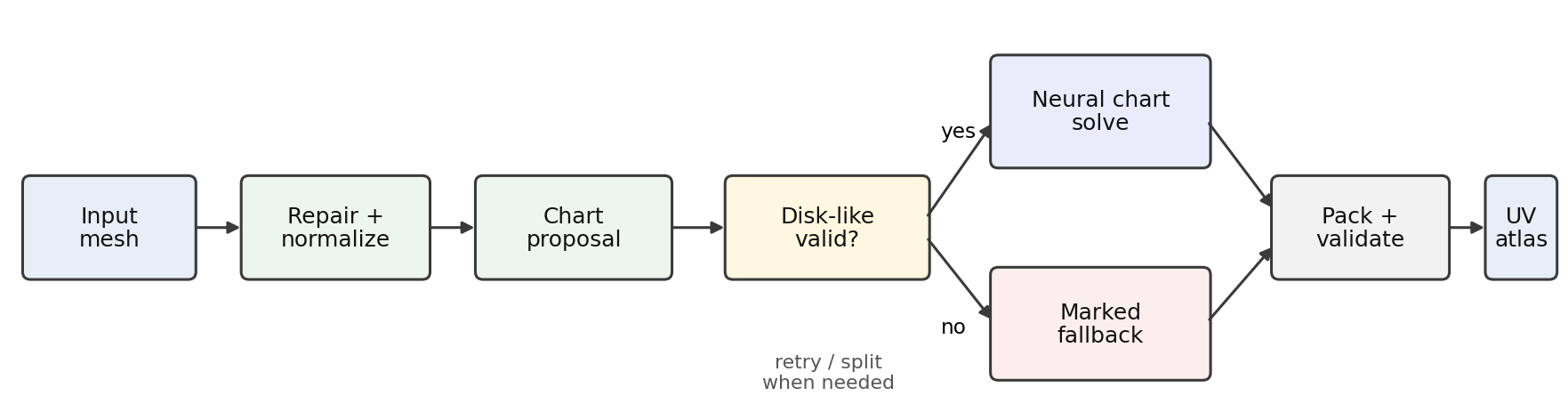}
    \caption{Validation-first atlas construction for generated meshes. The neural solver is invoked only on chart candidates that pass disk-like validation; rejected or retry-failed fragments are explicitly marked as fallback charts before packing and final atlas validation.}
    \label{fig:system_pipeline}
\end{figure*}

Algorithm~\ref{alg:pipeline} summarizes the chart-level solve, including feature construction, local-frame Jacobian assembly, Tutte fitting, and validity-checked optimization; Figure~\ref{fig:solver_schematic} diagrams the same forward path and validity loop. Figure~\ref{fig:system_pipeline} places this chart solver inside the generated-mesh routing and packing system.

\begin{algorithm}[!t]
\caption{Neural chart parameterization}
\label{alg:pipeline}
\small
\begin{algorithmic}[1]
\REQUIRE Disk-like chart $\cM=(V,T)$, spectral rank $k$, iterations $N$
\STATE Normalize vertices and build features $\mathbf{x}_i=[\tilde{\mathbf{p}}_i\|\tilde\psi_{1:k}(v_i)]$
\STATE Precompute local frames, areas $A_t$, and $\Delta C_t^{-1}$ for all $t\in T$
\STATE Compute a Tutte embedding $\mathbf{u}^{\mathrm{Tutte}}$ on a convex boundary
\STATE Initialize SIREN $f_\theta$ with a zero residual head and pretrain to $\mathbf{u}^{\mathrm{Tutte}}$
\FOR{$s=1$ to $N$}
    \STATE Predict $\mathbf{u}_i=f_\theta(\mathbf{x}_i)$ and assemble $J_t=\Delta U_t\Delta C_t^{-1}$
    \STATE Set $q(d)=d^{-2}$ if $d>\epsilon_J$, else use the matched $C^2$ Taylor polynomial
    \STATE Minimize $\sum_t A_t\|J_t\|_{\mathrm F}^2(1+q(\det J_t))+\alpha(s)\sum_t[\epsilon-\det J_t]_+^2$
    \STATE Save model/optimizer state, take an Adam step, and recompute determinants
    \IF{any triangle flips or generated-chart quality thresholds fail}
        \STATE Restore the previous state, reduce the learning rate, and retry or mark fallback
    \ENDIF
\ENDFOR
\RETURN Foldover-free chart UVs, or an explicitly marked fallback chart after retry failure
\end{algorithmic}
\end{algorithm}

\section{Experiments}
\label{sec:experiments}

\paragraph{Data and baselines.}
We evaluate\footnote{\url{https://github.com/01C-Amara/NeuralUVAtlas}} three compact pre-cut chart meshes (Hand, Bob, Camel), Thingi10K/xatlas-cut chart sweeps~\cite{zhou2016thingi10k,xatlas}, and a 25-asset Amara Spatial generated-mesh set~\cite{amara_spatial}. Compact-chart baselines are LSCM, ARAP, and SLIM through libigl 2.6.2~\cite{libigl}, Boundary First Flattening (BFF) through a Docker-built command-line v1.6 binary, and OptCuts as a recutting baseline; to avoid under-tuned SLIM comparisons, we reran SLIM up to 1{,}000 iterations and report the tuned rows below. For generated meshes we additionally compare Blender Smart UV Project, Angle Based, and Conformal unwrap using per-face-corner loop UVs. Metrics are true final $E_{SD}$, conformal distortion, area distortion, flipped-triangle percentage, minimum determinant, and runtime. Checkerboard visualizations use only a procedural diagnostic texture, not input texture images.

\begin{table}[!t]
\centering
\caption{Compact pre-cut chart benchmark. LSCM, ARAP, SLIM (1{,}000 iters), and BFF are fixed-chart solvers that preserve the supplied seams; OptCuts recuts and is listed for reference, since it solves an easier, different-topology problem. \textbf{Bold} is the best value among foldover-free fixed-chart solvers; the distortion of maps that still contain flips is greyed, as it is not a meaningful comparison. Dashes indicate initialization or cleanup failure. Runtimes are the full 5{,}000-iteration solve; a short validated pass reaches comparable quality in ${\sim}6$\,s (Fig.~\ref{fig:runtime_pareto}). Among fixed-chart solvers, only ours is foldover-free on every chart.}
\label{tab:compact_prelim}
\scriptsize
\setlength{\tabcolsep}{2pt}
\resizebox{\linewidth}{!}{%
\begin{tabular}{@{}llrrrrr@{}}
\toprule
Mesh & Method & $E_{SD}\!\downarrow$ & Conf.$\downarrow$ & Area$\downarrow$ & \%Flip$\downarrow$ & Time (s) \\
\midrule
Hand & LSCM & \inv{$1.47{\times}10^9$} & \inv{8.98} & \inv{27.71} & 13.56\% & 0.001 \\
Hand & ARAP & \inv{1262.61} & \inv{10.47} & \inv{1.58} & 15.74\% & 0.004 \\
Hand & SLIM-1k & \inv{$7.86{\times}10^7$} & \inv{7.85} & \inv{27.54} & 13.56\% & 1.194 \\
Hand & BFF & \inv{$6.20{\times}10^7$} & \inv{8.35} & \inv{27.59} & 13.80\% & 0.305 \\
Hand & Ours & \textbf{12.73} & \textbf{4.68} & \textbf{0.73} & \textbf{0.00\%} & 23.44 \\
\addlinespace[1.5pt]
Hand & OptCuts (re-cut) & 4.10 & 1.16 & 0.02 & 0.00\% & 1.755 \\
\midrule
Bob & LSCM & --- & --- & --- & --- & --- \\
Bob & ARAP & --- & --- & --- & --- & --- \\
Bob & SLIM-1k & --- & --- & --- & --- & --- \\
Bob & BFF & 14.04 & \textbf{1.24} & 1.08 & \textbf{0.00\%} & 0.305 \\
Bob & Ours & \textbf{4.39} & 1.30 & \textbf{0.07} & \textbf{0.00\%} & 37.45 \\
\addlinespace[1.5pt]
Bob & OptCuts (re-cut) & --- & --- & --- & --- & --- \\
\midrule
Camel & LSCM & \inv{30909.10} & \inv{1.16} & \inv{13.96} & 0.22\% & 0.007 \\
Camel & ARAP & \inv{$1.71{\times}10^8$} & \inv{406.22} & \inv{1.62} & 7.61\% & 0.035 \\
Camel & SLIM-1k & \textbf{4.09} & 1.14 & \textbf{0.02} & \textbf{0.00\%} & 6.381 \\
Camel & BFF & 4.27 & \textbf{1.10} & 0.17 & \textbf{0.00\%} & 0.348 \\
Camel & Ours & 4.14 & 1.18 & 0.07 & \textbf{0.00\%} & 27.80 \\
\addlinespace[1.5pt]
Camel & OptCuts (re-cut) & 4.09 & 1.15 & 0.02 & 0.00\% & 0.110 \\
\bottomrule
\end{tabular}}
\end{table}

Table~\ref{tab:compact_prelim} should be read as a failure-mode diagnostic, not as a claim that classical solvers are generally weak. With 1{,}000 iterations, SLIM improves Camel to $E_{SD}=4.09$ with zero flips and remains faster than our PyTorch solver; on this chart, SLIM is the better fixed-chart solver. BFF is also strong on clean or multi-component inputs: it solves Bob with zero flips and low conformal distortion, though at higher $E_{SD}$ and area distortion than ours. Most importantly, OptCuts resolves Hand and Camel when recutting is allowed, reaching near-isometric zero-flip maps quickly. Its Bob cleanup failure reflects a stricter recutting/preprocessing contract on this disconnected input, while BFF's loop-UV export and our chart solver can still handle the separated components. This narrows the compact-chart claim: our neural solver is not a replacement for joint cut optimization, but a fixed chart-level solver that remains useful when preserving a supplied chart or when recutting/cleanup fails, as on Bob for libigl and OptCuts. Table~\ref{tab:slim_failure_modes} summarizes the rerun.

\begin{table}[!t]
\centering
\caption{Compact baseline failure-mode check. $C/B$ reports connected components and boundary loops from the topology scan; SLIM, BFF, and OptCuts entries show flip rate except for valid rows, where $E_{SD}$ is more diagnostic. OptCuts is allowed to recut.}
\label{tab:slim_failure_modes}
\scriptsize
\setlength{\tabcolsep}{3pt}
\begin{tabular}{@{}lccccc@{}}
\toprule
Mesh & $C/B$ & SLIM-20 & SLIM-1k & BFF & OptCuts \\
\midrule
Hand & 1/1 & 13.56\% flips & 13.56\% flips & 13.80\% flips & $E_{SD}=4.10$ \\
Bob & 3/3 & init fail & init fail & $E_{SD}=14.04$ & cleanup fail \\
Camel & 1/1 & $E_{SD}=4.13$ & $E_{SD}=4.09$ & $E_{SD}=4.27$ & $E_{SD}=4.09$ \\
\bottomrule
\end{tabular}
\end{table}

\begin{figure}[!t]
    \centering
    \includegraphics[width=0.98\linewidth]{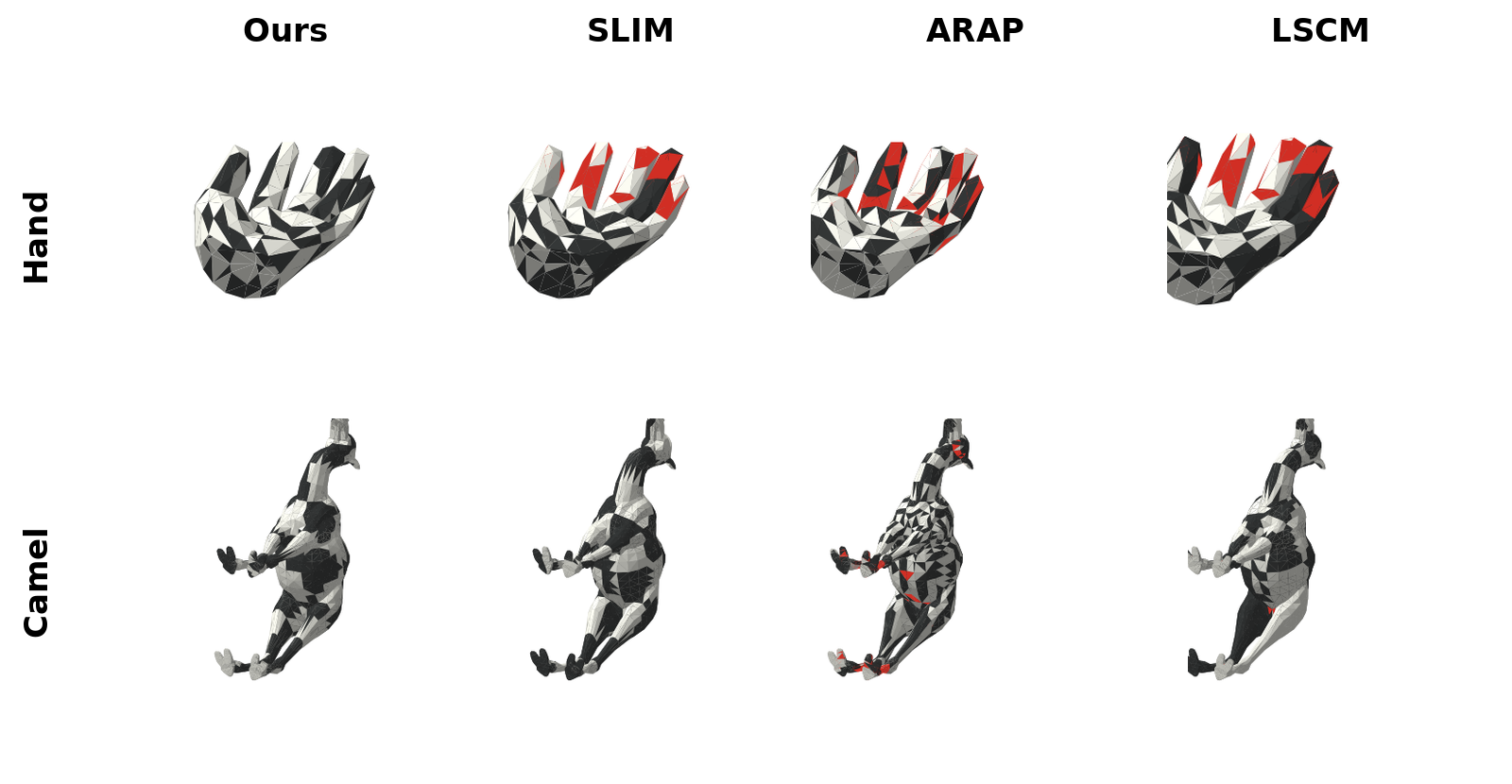}
    \vspace{0.1em}
    \includegraphics[width=0.98\linewidth]{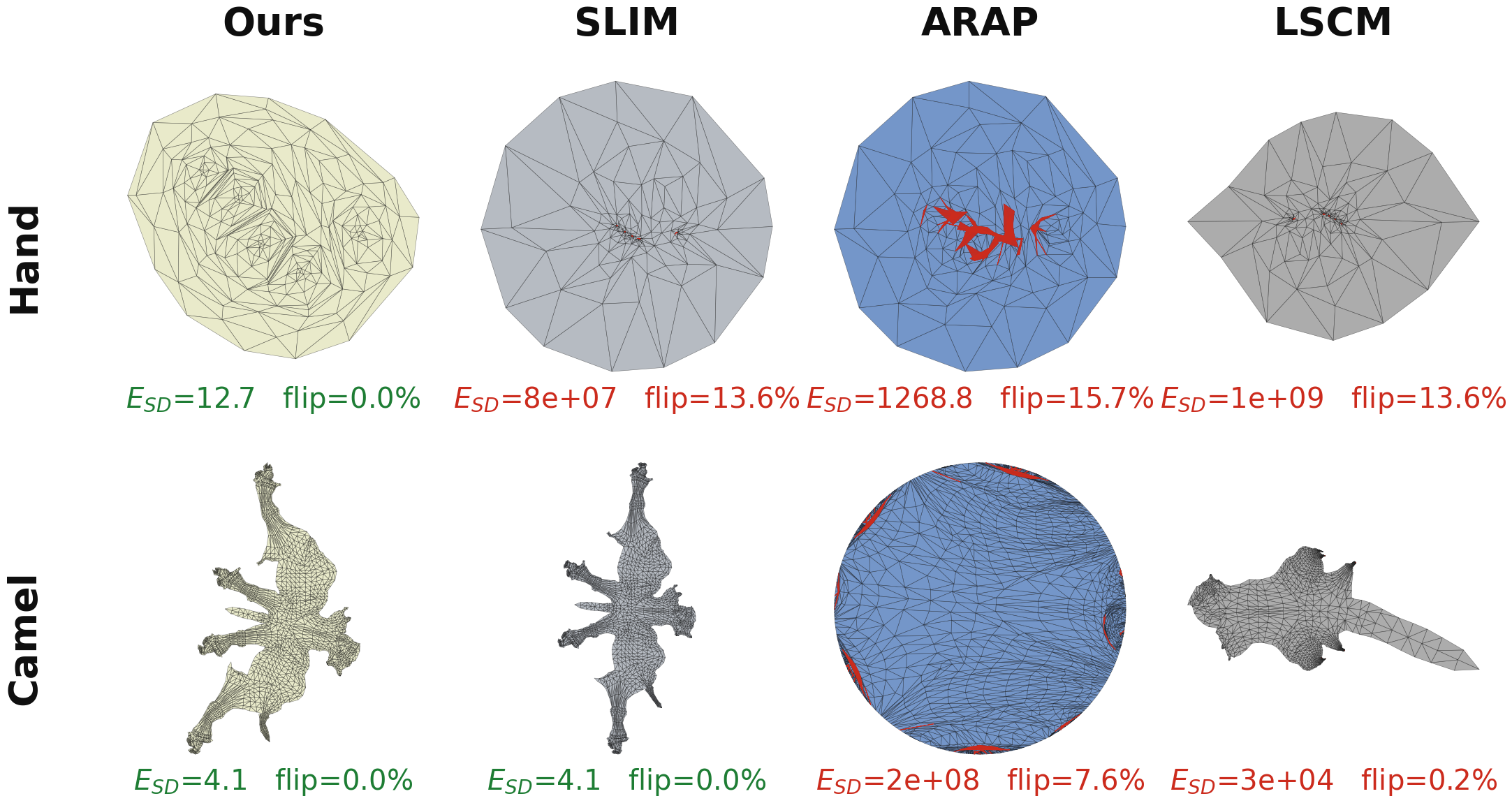}
    \caption{Compact benchmark maps for the representative single-component charts. Top: checkerboard transfer; bottom: UV layouts. Red marks flipped triangles; Bob is omitted from this plate because the libigl baselines fail on its three-component input.}
    \label{fig:compact_maps}
\end{figure}

\begin{figure}[!t]
    \centering
    \includegraphics[width=0.98\linewidth]{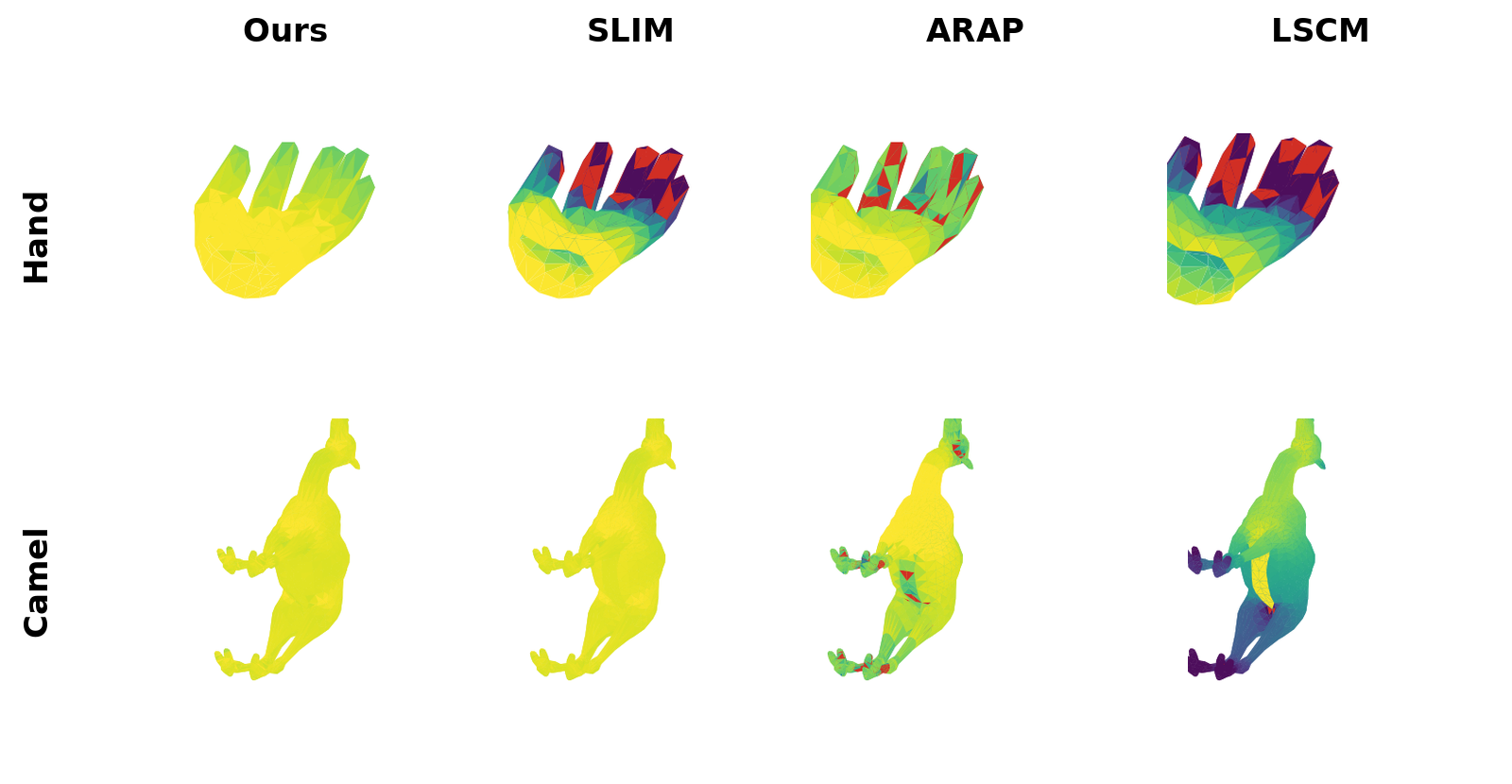}
    \vspace{0.1em}
    \includegraphics[width=0.98\linewidth]{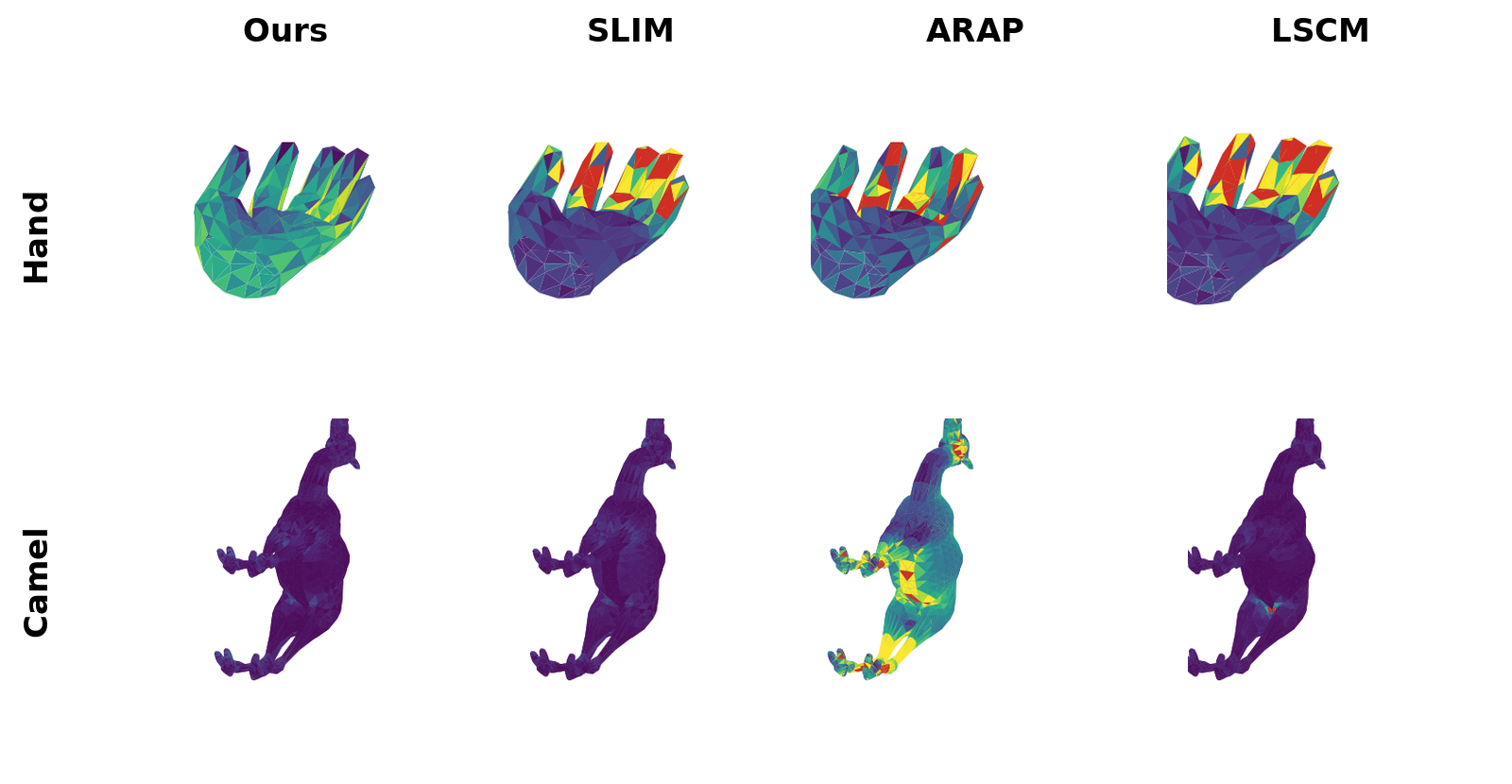}
    \caption{Compact benchmark heatmaps. Top: Jacobian determinant; bottom: conformal distortion. Red/non-positive determinant regions expose invalid direct-solver maps.}
    \label{fig:compact_heatmaps}
\end{figure}

\begin{figure}[!t]
    \centering
    \includegraphics[width=\linewidth]{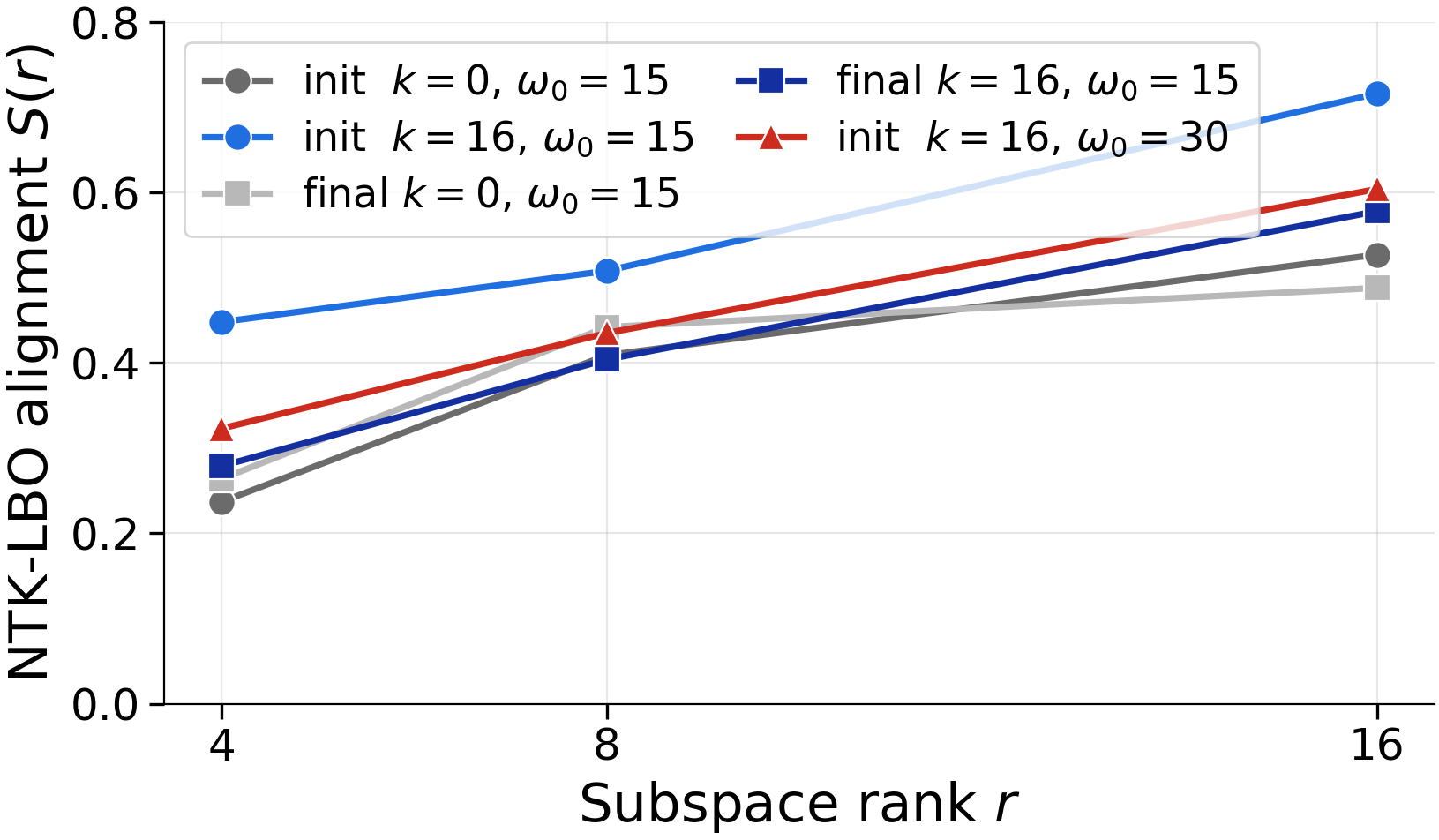}
    \caption{Compact NTK--LBO alignment curves averaged over Hand, Bob, and Camel. Spectral features raise alignment most clearly at initialization; the high-bandwidth $\omega_0=30$ setting weakens this alignment.}
    \label{fig:ntk_compact_controls}
\end{figure}

\paragraph{Ablations and visuals.}
Table~\ref{tab:ablation_prelim} separates the roles of spectral inputs, SIREN bandwidth, and warm-up. Removing LBO features hurts distortion and determinant margin but does not by itself introduce flips on the compact charts; the bandwidth rows are nearly flat, so we treat $\omega_0$ as a capacity setting rather than the central prior. At the 1{,}000-iteration budget on all 47 stratified Thingi charts, the same $k=0$ versus $k=16$ check gives a different scale behavior: $k=16$ solves one additional chart validly (42/47 vs.\ 41/47), but has higher median $E_{SD}$ (99.68 vs.\ 41.66) and lower median determinant margin (0.047 vs.\ 0.067). On larger charts, spectral inputs are therefore best read as a validity-oriented trade-off at this budget: they recover one additional zero-flip solve, but pay a distortion and margin cost. The hard validity dependence is the Tutte/residual initialization plus barrier, as random or spectral absolute initializations become invalid. Figures~\ref{fig:compact_maps} and~\ref{fig:compact_heatmaps} show the corresponding map and heatmap diagnostics. Table~\ref{tab:ntk_compact_controls} and Figure~\ref{fig:ntk_compact_controls} report NTK--LBO controls averaged over Hand, Bob, and Camel, followed by the larger Amara fragment probe.

\begin{table}[!t]
\centering
\caption{Compact ablations averaged over Hand, Bob, and Camel. Bandwidth rows keep $k=16$ and vary only $\omega_0$; their flatness indicates that $\omega_0$ is not the dominant validity factor in this regime.}
\label{tab:ablation_prelim}
\scriptsize
\setlength{\tabcolsep}{3pt}
\begin{tabular}{@{}lccc@{}}
\toprule
Variant & $E_{SD}$ & \%Flip & $\min\det J$ \\
\midrule
Full ($\omega_0=15,k=16$) & 7.09 & 0.00\% & 0.27 \\
No spectral ($k=0$) & 8.01 & 0.00\% & 0.19 \\
$k = 8$ & 7.09 & 0.00\% & 0.24 \\
$k = 32$ & 7.08 & 0.00\% & 0.27 \\
$\omega_0=5$ & 7.10 & 0.00\% & 0.26 \\
$\omega_0=10$ & 7.09 & 0.00\% & 0.22 \\
$\omega_0=20$ & 7.08 & 0.00\% & 0.27 \\
$\omega_0=30$ & 7.09 & 0.00\% & 0.23 \\
Spectral init & invalid & 49.70\% & -11.78 \\
Random init & invalid & 50.60\% & -33.47 \\
\bottomrule
\end{tabular}
\end{table}

\begin{table}[!t]
\centering
\caption{NTK--LBO subspace alignment. Compact rows average Hand, Bob, and Camel; Amara rows use 100 accepted and 100 fallback-routed generated-mesh fragments. Higher $S(r)=\|Q_r^\top\Psi_r\|_{\mathrm F}^2/r$ means stronger alignment with the first $r$ LBO modes.}
\label{tab:ntk_compact_controls}
\scriptsize
\setlength{\tabcolsep}{3pt}
\begin{tabular}{@{}llrrrrr@{}}
\toprule
Set & State & $k$ & $\omega_0$ & $S(4)$ & $S(8)$ & $S(16)$ \\
\midrule
Compact & init & 0 & 15 & 0.237 & 0.409 & 0.527 \\
Compact & init & 16 & 15 & 0.448 & 0.508 & 0.716 \\
Compact & final & 0 & 15 & 0.264 & 0.442 & 0.488 \\
Compact & final & 16 & 15 & 0.279 & 0.404 & 0.578 \\
Compact & init & 16 & 30 & 0.323 & 0.435 & 0.604 \\
Amara acc. & init & 0 & 15 & 0.601 & 0.732 & 0.801 \\
Amara acc. & init & 16 & 15 & 0.505 & 0.611 & 0.905 \\
Amara fall. & init & 0 & 15 & 0.551 & 0.687 & 0.777 \\
Amara fall. & init & 16 & 15 & 0.502 & 0.610 & 0.906 \\
\bottomrule
\end{tabular}
\end{table}

\begin{figure}[!t]
    \centering
    \includegraphics[width=\linewidth]{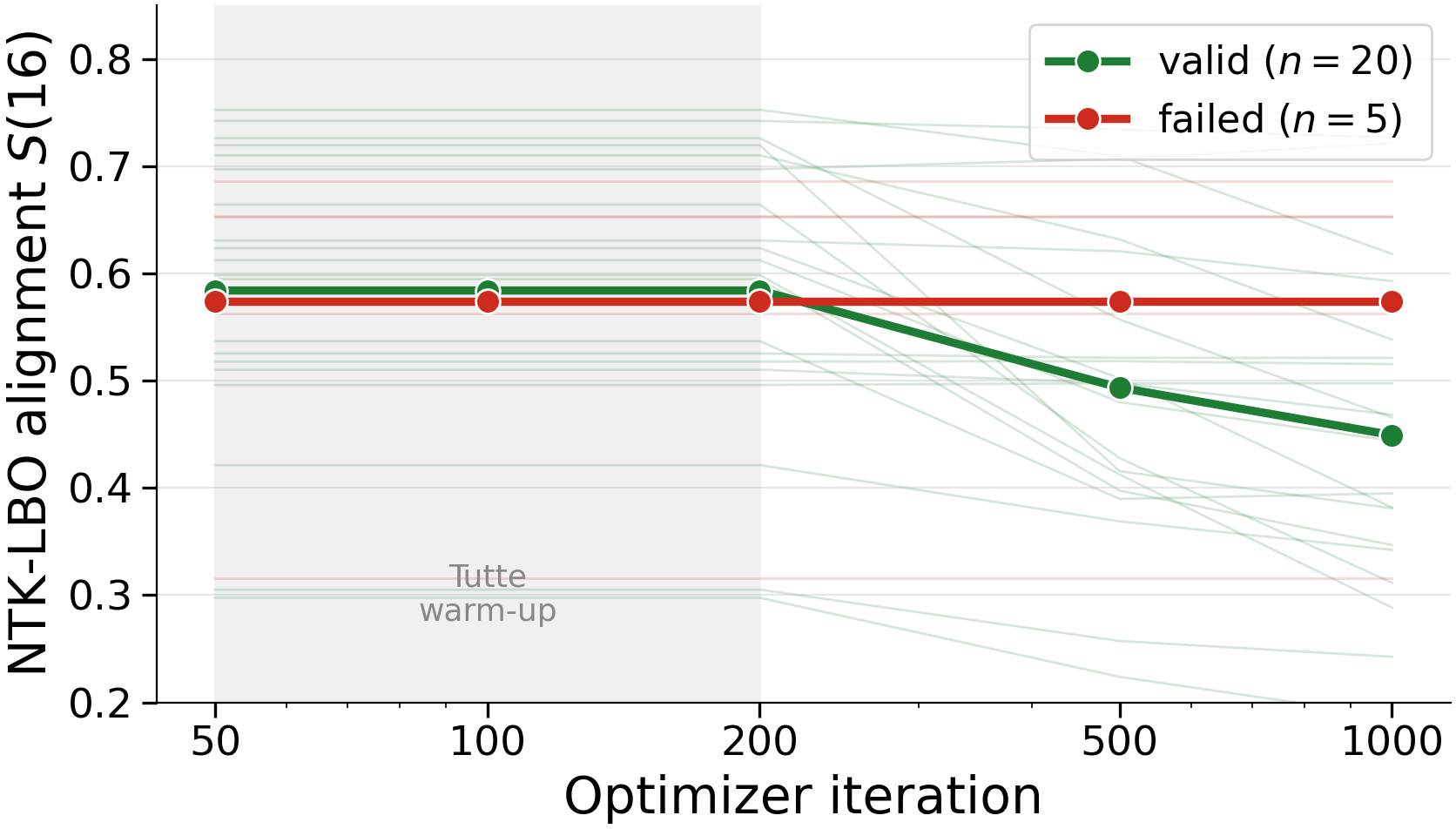}
    \caption{NTK--LBO alignment $S(16)$ over optimization, on the 25-chart Thingi trajectory probe (thin lines: individual charts; bold: group means). The Tutte warm-up is stationary; once energy optimization begins, charts that end valid drift away from the initial NTK--LBO subspace, while charts that end invalid stay stuck under flip-rejected steps. Alignment correlates only moderately with the outcome, so it is a mechanistic diagnostic, not a validity predictor.}
    \label{fig:ntk_trajectory}
\end{figure}

The NTK probes in Table~\ref{tab:ntk_compact_controls} are intentionally diagnostic rather than celebratory. On compact charts, spectral inputs raise alignment at initialization but the gap shrinks after optimization, suggesting that LBO conditioning acts mainly as a starting bias and the network later adapts to the distortion objective. A 25-chart Thingi trajectory probe (Fig.~\ref{fig:ntk_trajectory}) confirms this interpretation: the residual Tutte warm-up is stationary, successful charts often move away from the initial NTK--LBO subspace after energy optimization begins, and failed charts tend to remain stuck under flip-rejected steps. Early $S(r)$ values show moderate predictive signal in this probe, with strongest absolute correlations of $|r|=0.41$ with final flip rate and $|r|=0.44$ with final determinant margin over the measured features and milestones, but this is not strong enough to drive go/no-go validity decisions. On the 200-fragment Amara probe, accepted and fallback-routed fragments have nearly identical $S(16)$ after conditioning, so alignment is not an acceptance predictor. The consistent effect is more specific: LBO features raise $S(16)$ on 198/200 fragments (means $0.801$ to $0.905$ for accepted fragments and $0.777$ to $0.906$ for fallback-routed fragments), while lowering $S(4)$ and $S(8)$ on average. One interpretation is that spectral inputs move update mass from the lowest few modes into ranks 5--16, changing the available smooth-but-not-constant deformation directions without deciding chart validity on their own.

\begin{figure}[!t]
    \centering
    \includegraphics[width=\linewidth]{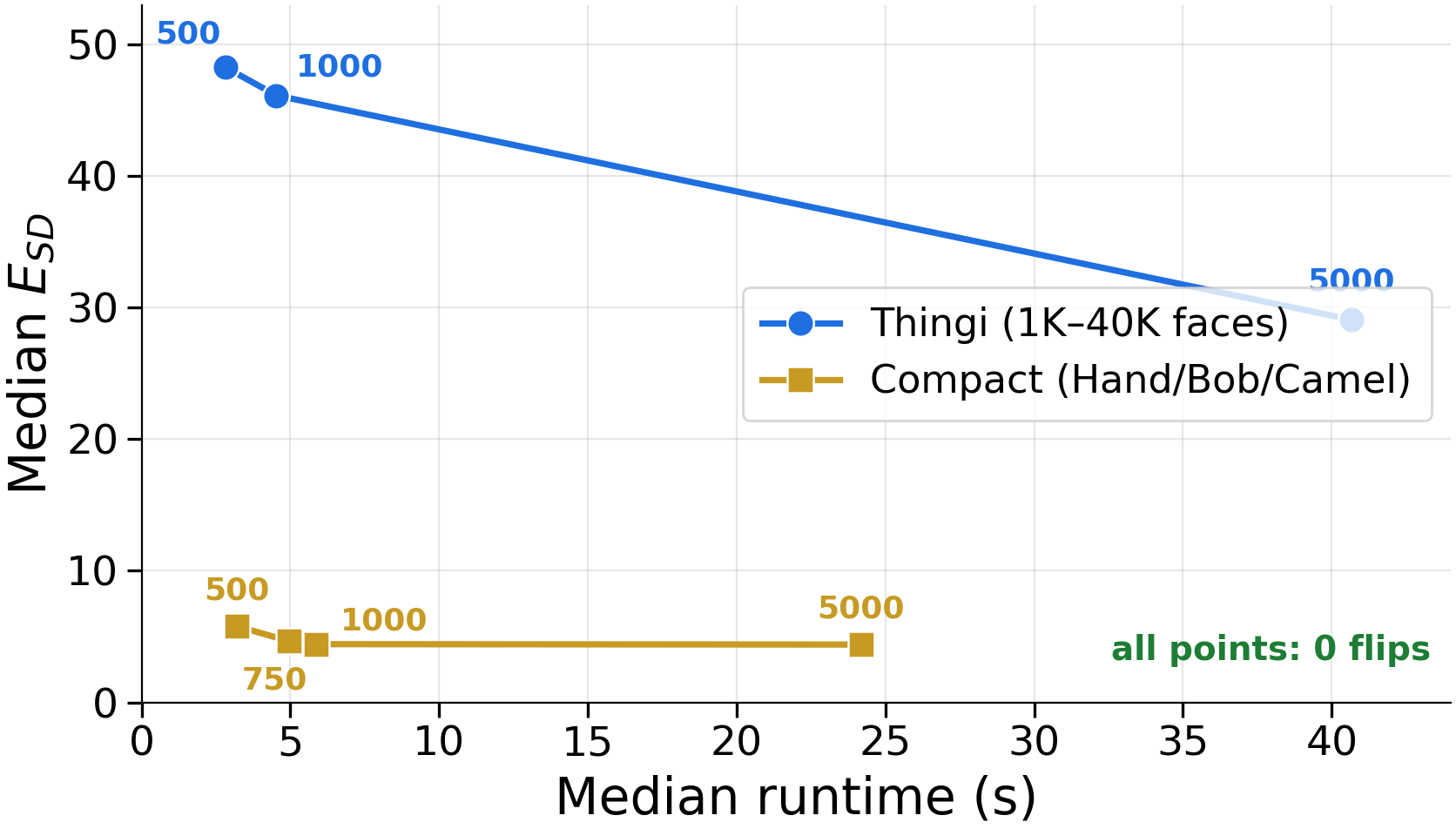}
    \caption{Runtime--quality trade-off from Table~\ref{tab:runtime_pareto} (median $E_{SD}$ vs.\ median runtime; iteration budget annotated at each point). Every plotted point is zero-flip. Compact-chart distortion saturates by $\sim$1{,}000 iterations, so a short validated pass recovers most of the quality, while larger Thingi charts keep improving with more iterations.}
    \label{fig:runtime_pareto}
\end{figure}

\paragraph{Scale and runtime.}
Table~\ref{tab:thingi_scale} reports all attempted charts rather than only clean successes. The strongest scale result is the small-chart xatlas first-500 sweep: 280/287 attempted charts are finite, positive-determinant, and zero-flip. The stratified xatlas-cut benchmark attempted 47 larger charts; five retry-confirmed failures remained flipped over seeds 0, 1, and 2, with best flip rates from 0.0097\% to 9.30\%. Table~\ref{tab:runtime_pareto} and Figure~\ref{fig:runtime_pareto} give a fixed-budget CUDA Pareto diagnostic. On compact charts, all tested budgets remain zero-flip and median distortion is nearly saturated by 750--1{,}000 iterations. A three-chart Thingi probe at roughly 1K, 6.5K, and 40K faces also stays zero-flip at 500 and 1{,}000 iterations, but its distortion continues improving at 5{,}000 iterations. As additional classical checks on the same 1K--40K-face Thingi sample, BFF solves 8/10 charts validly and OptCuts solves 9/10 validly when recutting is allowed; the remaining invalid cases have small flip rates below 0.05\%. This supports a practical atlas-construction strategy: run fast classical or recutting solvers when their assumptions are acceptable, use a short validated neural pass when supplied-chart validity is the requirement, and reserve longer optimization for larger or quality-critical charts.

\begin{table}[!t]
\centering
\caption{Thingi10K/xatlas-cut scale results with explicit denominators. Valid means finite UVs, zero flipped triangles, and positive final minimum determinant.}
\label{tab:thingi_scale}
\scriptsize
\setlength{\tabcolsep}{2pt}
\resizebox{\linewidth}{!}{%
\begin{tabular}{@{}lrrrrrr@{}}
\toprule
Subset & Tried & Finite & Valid & Face range & Med. $E_{SD}$ & Mean time \\
\midrule
Raw usable & 26 & 18 & 18 (69.2\%) & 165--131{,}072 & 5.27 & 36.29s \\
Xatlas first-500 & 287 & 283 & 280 (97.6\%) & 4--236 & 5.45 & 19.65s \\
Xatlas stratified & 47 & 47 & 42 (89.4\%) & 1{,}028--196{,}608 & 60.56 & 48.72s \\
\bottomrule
\end{tabular}}
\end{table}

\begin{table}[!t]
\centering
\caption{Runtime--quality Pareto sweep on CUDA. Compact rows average Hand, Bob, and Camel; Thingi rows use three valid stratified xatlas-cut charts at roughly 1K, 6.5K, and 40K faces. All rows use 200 warm-up iterations.}
\label{tab:runtime_pareto}
\scriptsize
\setlength{\tabcolsep}{3pt}
\begin{tabular}{@{}lccccc@{}}
\toprule
Set / iters & Valid & Med. $E_{SD}$ & Mean $\min\det J$ & Mean flip & Med. time \\
\midrule
Compact / 500 & 3/3 & 5.78 & 0.222 & 0.00\% & 3.21s \\
Compact / 750 & 3/3 & 4.65 & 0.206 & 0.00\% & 4.96s \\
Compact / 1{,}000 & 3/3 & 4.42 & 0.228 & 0.00\% & 5.88s \\
Compact / 5{,}000 & 3/3 & 4.39 & 0.256 & 0.00\% & 24.18s \\
Thingi / 500 & 3/3 & 48.28 & 0.214 & 0.00\% & 2.82s \\
Thingi / 1{,}000 & 3/3 & 46.14 & 0.228 & 0.00\% & 4.53s \\
Thingi / 5{,}000 & 3/3 & 29.09 & 0.315 & 0.00\% & 40.66s \\
\bottomrule
\end{tabular}
\end{table}

\begin{table}[!t]
\centering
\caption{Amara coverage and validity. Ours is measured on 1{,}219{,}074 repaired faces; Blender denominators use 1{,}219{,}095 raw loop-UV triangles.}
\label{tab:amara_coverage_validity}
\scriptsize
\setlength{\tabcolsep}{2pt}
\begin{tabular}{@{}lccc@{}}
\toprule
Method & Coverage & Routed faces & Validity / flips \\
\midrule
Ours neural-only & 52.1\% & neural & 4{,}708/4{,}708 charts \\
Ours full fallback & 100.0\% & 52.1/47.9\% & 0 local flips \\
Blender Smart UV & 100.0\% & n/a & 246/1{,}219{,}095 (0.020\%) \\
\bottomrule
\end{tabular}
\end{table}

\begin{table}[!t]
\centering
\caption{Blender chart-level baselines on Amara.}
\label{tab:amara_blender_methods}
\scriptsize
\setlength{\tabcolsep}{2pt}
\begin{tabular}{@{}lcccc@{}}
\toprule
Method & OK & Zero-flip & Flips & Conf. \\
\midrule
Angle Based & 25/25 & 16/25 & 162 & 1.01 \\
Conformal & 25/25 & 13/25 & 146 & 1.01 \\
Smart UV & 25/25 & 14/25 & 246 & 1.23 \\
\bottomrule
\end{tabular}
\end{table}

\begin{table}[!t]
\centering
\caption{Larger local Amara atlas runs with the Rust implementation. Rows are deterministic path-order prefixes of the 10{,}071 local GLB assets; the 1K row combines the initial run with targeted timeout retry and fallback writeback for failed packed islands. These runs test validation, routing, and packing, with optional bounded neural optimization only in the paper-native row.}
\label{tab:amara_large_sweep}
\scriptsize
\setlength{\tabcolsep}{2pt}
\resizebox{\linewidth}{!}{%
\begin{tabular}{@{}lrrrrrrr@{}}
\toprule
Run & Tried & OK & Faces & Flips & Neural-routed/fallback & Opt. charts & Med. time \\
\midrule
Auto, 1K + retry & 1{,}000 & 1{,}000 & 48.5M & 0 & 60.8/39.2\% & 0 & 11.37s \\
Paper-native, 100 & 100 & 100 & 4.85M & 0 & 66.9/33.1\% & 38 & 8.26s \\
\bottomrule
\end{tabular}}
\end{table}

\begin{figure}[!t]
    \centering
    \includegraphics[width=\linewidth]{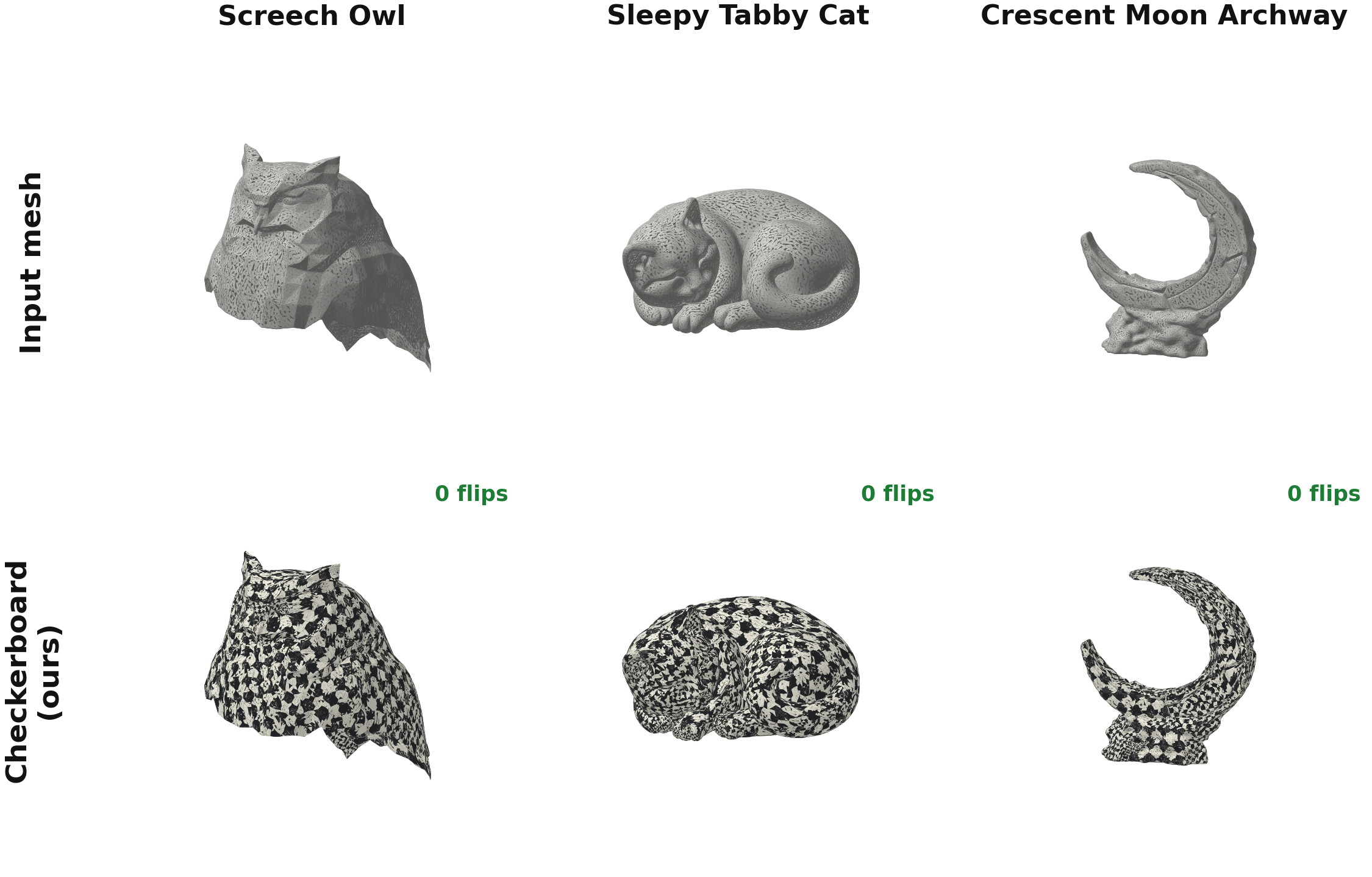}
    \caption{Qualitative atlases on Amara Spatial generated meshes. Top: input mesh; bottom: checkerboard transfer through our atlas (neural charts plus marked fallbacks). All transfers are foldover-free across these varied generated shapes.}
    \label{fig:amara_gallery}
\end{figure}

\begin{figure*}[!t]
    \centering
    \includegraphics[width=\textwidth]{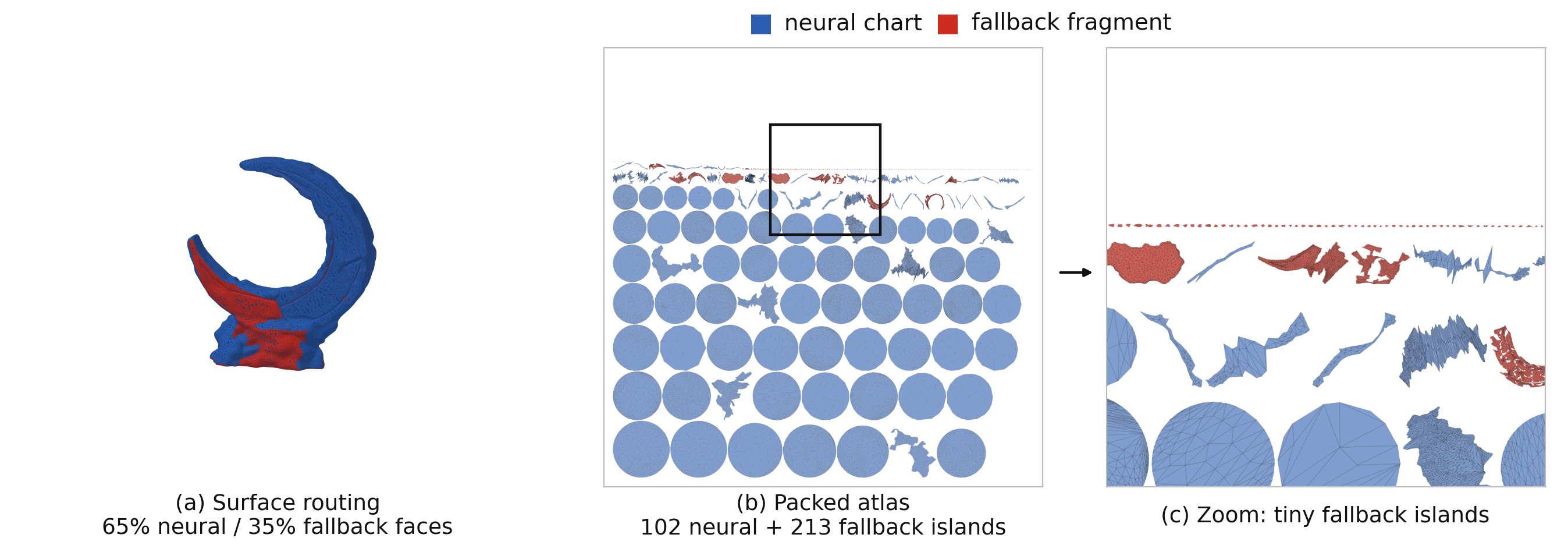}
    \caption{Generated-mesh routing and its limitation, on a Crescent Moon Archway asset. (a) The repaired surface colored by routing decision: disk-like regions are accepted as neural charts (blue), while thin or non-disk fragments are routed to marked fallbacks (red). (b) The packed atlas under the same coloring: a few large neural islands plus a band of fallbacks. (c) Zooming the fallback band shows the proliferation of tiny islands. The atlas is locally valid and complete, but messy generated fragments create many tiny islands, motivating better chart merging, recutting, remeshing, and packing.}
    \label{fig:amara_fallback_limitation}
\end{figure*}

\paragraph{Generated-mesh study.}
The 25-asset Amara Spatial study evaluates the solver inside a conservative raw-mesh atlas-construction setting. The central finding is a routing split: only 52.1\% of repaired faces are accepted as disk-like neural charts, while 47.9\% are generated fragments that violate the chart-quality assumptions and are explicitly routed to PCA or per-face fallback. The accepted neural portion contains 4{,}708 charts and all are solved with zero flips. The atlas construction path covers 1{,}219{,}074 repaired faces with zero local flipped triangles; Table~\ref{tab:amara_coverage_validity} summarizes these coverage and validity results. Blender Smart UV Project covers the full assets and is valid on 99.98\% of raw loop-UV triangles, but still produces 246 flipped UV triangles across 11/25 assets; Table~\ref{tab:amara_blender_methods} reports the corresponding Blender chart-level baselines. The repaired-mesh coverage count uses 1{,}219{,}074 faces, while Blender flip counts use the raw denominator of 1{,}219{,}095 loop-UV triangles. Figure~\ref{fig:amara_gallery} shows foldover-free checkerboard transfers on several generated assets, and Figure~\ref{fig:amara_fallback_limitation} illustrates the routing split and the resulting tiny fallback islands.

\paragraph{Larger Amara sweep.}
To test scale beyond the 25-asset study, we ran deterministic prefixes of the 10{,}071 local Amara GLB assets through the Rust atlas implementation. The auto pass processed 1{,}000/1{,}000 assets successfully after serially rerunning the initial timeout cases and routing failed packed islands to a deterministic per-face writeback fallback, covering 48.5M successful input faces with zero UV flips. This row measures atlas validation, routing, writeback, and packing, not neural optimization. A bounded paper-native pass on the same first 100 assets optimized 38 neural charts and also produced zero flips over 4.85M faces. Table~\ref{tab:amara_large_sweep} reports these larger sweeps. These results support the routing-and-validation study, but they should not be read as full PyTorch neural solves for every chart.

\section{Discussion and Conclusion}
\label{sec:conclusion}

The experiments point to a hybrid computational strategy. The full PyTorch neural chart solve is slower than classical solvers, but the fixed-budget Pareto sweep shows that a short validated pass can recover most compact-chart quality: 750--1{,}000 iterations remain zero-flip and approach the 5{,}000-iteration distortion in under six median seconds on the test GPU. The Thingi probe shows the complementary scale behavior: short passes preserve validity on representative larger charts, while longer optimization can still improve distortion. In practice, direct or recutting solvers can be used when their assumptions are acceptable, a short neural pass can be used when supplied-chart validity is the requirement, and longer optimization can be reserved for large or stubborn cases. Table~\ref{tab:amara_large_sweep} supports the validation and packing part of this strategy at larger scale: the Rust atlas implementation validates, routes, writes, and packs generated assets with zero local flips after fallback routing, while bounded paper-native optimization is reserved for selected small charts. These large-scale routing numbers are not substitutes for the full PyTorch chart solver, but they show how the solver can sit inside validation-first atlas construction.

The Amara result also gives a failure-mode taxonomy for generated meshes. The solver assumes disk-like charts; arbitrary generated fragments still require robust repair, seam proposal, retry/splitting, and high-quality packing. In the 25-asset study, 52.1\% of repaired faces were accepted for neural solving, while 47.9\% were routed to marked fallbacks. Our fallback path gives local validity and coverage, not a claim that every invalid fragment has received a high-quality neural parameterization.

The closest learned-atlas systems remain important future comparisons. Nuvo and Neural Jacobian Fields target whole-surface neural atlases or deformation/Jacobian fields rather than this paper's fixed chart-level objective, so a fair comparison requires an adapter that exports loop-level UVs and evaluates the same final distortion and flip metrics. We found no official Nuvo code or checkpoint release linked from the paper or project page. The official NJF repository provides morphing checkpoints and a UV code path, but the visible UV script references an unavailable trained UV checkpoint and author-local datasets. Local feasibility probes with an unofficial Nuvo implementation required either a fixed-triangle adapter or a seam-split export that changes the evaluated face set, so we avoid reporting non-comparable numbers; adding official neural-UV comparisons and progressive-embedding baselines is the most direct way to broaden the empirical comparison.

We have presented a continuous neural reparameterization framework for UV unwrapping. By optimizing an untrained SIREN conditioned on spectral mesh features, and by pairing it with Tutte warm-up plus a stable injectivity-aware objective, the method adds a practical solver recipe to classical distortion minimization. The denominator-aware Thingi10K and Amara results show where that recipe is useful, while the NTK--LBO study isolates what spectral conditioning does, namely that it changes update geometry, from what determines chart validity, which remains initialization, barriers, chart quality, and routing. The clearest next steps are runtime reduction, automatic chart construction, external neural baselines, and atlas packing.

\section*{Acknowledgments}
The author thanks the Zero One Creative team for helpful discussions, and Raymond Wong in particular for insights into the implementation.
{\small
\bibliographystyle{ieeenat_fullname}

\end{document}